\documentstyle[12pt]{article}

\newcommand{\be}{\begin{equation}}
\newcommand{\ee}{\end{equation}}
\newcommand{\ba}{\begin{eqnarray}}
\newcommand{\ea}{\end{eqnarray}}
\begin{document}
\def\pct#1{(see Fig. #1.)}

\begin{titlepage}
\hbox{\hskip 12cm CERN-TH/96-180  \hfil}
\hbox{\hskip 12cm ROM2F-96/40  \hfil}
\hbox{\hskip 12cm UCLA/96/TEP/22  \hfil}
\hbox{\hskip 12cm hep-th/9607105 \hfil}
\begin{center}  {\Large  \bf  Twelve-Dimensional \ Aspects of }
\vskip 18pt {\Large  \bf  Four-Dimensional \ N=1 \ Type I \ Vacua}
 
\vspace{1.8cm}
 
{\large \large M. Bianchi$^{*}$, S. Ferrara$^{\dagger}$, G. Pradisi$^{*}$, A.
Sagnotti$^{*}$ and Ya.S. Stanev$^{*}$\footnote{I.N.F.N.   Fellow, on Leave from
Institute for Nuclear Research and Nuclear Energy, Bulgarian  Academy of Sciences,
BG-1784 Sofia, BULGARIA.}}

\vspace{0.7cm}

{$^{\dagger}$ \sl Theory \ Division, \ \ CERN  \\ CH-1211  Geneva 23, \ \  SWITZERLAND}

\vspace{0.7cm}

{$^{*}$ \sl Dipartimento di Fisica, \ \ Universit{\`a} di Roma \ ``Tor Vergata''
\\ I.N.F.N.\ - \ Sezione di Roma \ ``Tor Vergata'', \ \ Via della Ricerca Scientifica ,
1
\\ 00133 \ Roma, \ \ ITALY}
\end{center}
\vskip 1cm

\abstract{Four-dimensional supergravity theories are reinterpreted in a 12-dimensional
F-theory framework. The $O(8)$ symmetry of $N=8$ supergravity is related to a reduction
of F-theory on $T_8$, with the  seventy scalars formally associated, by $O(8)$
triality, to a fully compactified four-form $A_4$. For the
$N=1$ type I model recently obtained from the type IIB string on the $Z$ orbifold, we
identify the K\"ahler manifold of the untwisted scalars in the unoriented closed sector
with the generalized Siegel upper-half plane
$Sp(8,R)/(SU(4) \times U(1))$. The $SU(4)$ factor reflects the holonomy group of
Calabi-Yau fourfolds.}
\vskip 24pt
\hbox{\hskip 1.2cm CERN-TH/96-180  \hfil}
\hbox{\hskip 1.2cm July  1996 \hfil}
\vfill \end{titlepage}
\makeatletter
\@addtoreset{equation}{section}
\makeatother
\renewcommand{\theequation}{\thesection.\arabic{equation}}
\addtolength{\baselineskip}{0.3\baselineskip} 
\section{Introduction}

Non-trivial backgrounds for the scalar fields of type IIB supergravity have provided a
geometrical setting \cite{vafaf,mv} for some peculiar six-dimensional (6D) string vacua
\cite{bs} previously derived as open descendants
\cite{cargese} of type IIB K3 compactifications. The resulting models differ markedly 
from conventional K3 reductions, since their massless spectra contain different numbers
of (anti)self-dual tensors that take part in a generalized Green-Schwarz mechanism
\cite{tensor,fms}. These peculiar features find an elegant rationale in the
compactification of a putative 12D F-theory \cite{vafaf} on elliptically fibered
Calabi-Yau (CY) threefolds 
\cite{mv}, a construction that generalizes previous work on supergravity vacua with
scalar backgrounds \cite{opgz} by taking into account subtle global issues.  Many of
the recent developments can be traced to some remarkable properties of the scalar
manifolds of supergravity theories that have been known since the late seventies
\cite{csf}. These non-compact spaces admit a natural action of generalized dualities, 
and their negative curvatures allow for non-trivial scalar  backgrounds compatibly with
supersymmetry and vanishing vacuum energy.

In this letter we consider the possibility of endowing 4D supergravities with a  12D
interpretation in the spirit of F-theory. This relates the internal part of the 12D
Lorentz group to the
$O(8)$ symmetry of $N=8$ supergravity \cite{cj,dwn}, only an 
$O(7)$ subgroup of which is manifest in the 11D interpretation \cite{cjs}. In CY
compactifications of 10D superstrings, an
$SU(3)$ subgroup of the internal $O(6)$ is identified with the holonomy of the CY
threefold. In a similar fashion, in the decomposition of 
$O(8)$ into $O(2) \times O(6)$, $SO(8)$ triality allows a natural interpretation of the
$O(6)\sim SU(4)$ subgroup as the holonomy of a CY fourfold. The chiral type I vacua
recently constructed in
\cite{wroma2} offer a nice setting to elicit this correspondence. Just as the
``parent'' type IIB string on the $Z$ orbifold
\cite{dhvw} may be regarded as a singular limit of a compactification  on a CY
threefold, these $N=1$ models may be regarded as singular limits of F-theory
compactifications on CY fourfolds, some examples of which have been recently discussed
in \cite{cy4}. In the following we give evidence that the  K\"ahler manifold of the
untwisted scalars in the projected theory is
$Sp(8,R)/(SU(4) \times U(1))$, precisely in the spirit of this correspondence.

\vskip 24pt
\section{The Scalar Manifold}

Let us begin by reviewing the results in \cite{wroma2}, where open descendants  of the
$Z$ orbifold limit of a CY threefold have been constructed. The unoriented
closed-string sector is obtained combining the torus amplitude ${\cal T}$ for the type
IIB string on the $Z$ orbifold \cite{dhvw} with a Klein-bottle projection ${\cal K}$.  
The original massless closed spectrum containing the $N=2$ supergravity multiplet,
$(9+1)$ hypermultiplets from the untwisted sector and 27 additional hypermultiplets
from the twisted sectors is thus truncated to $N=1$ supergravity coupled to $(1 + 9 +
27)$ chiral multiplets. The tadpole cancellation conditions require the  introduction
of open-string states with a Chan-Paton gauge group 
$SO(8) \times SU(12) \times U(1)$, and the resulting massless spectrum  includes three
generations  of chiral multiplets in the $(8,12^*)$ and $( 1, 66)$ representations. The
$U(1)$ factor is anomalous, and  a generalized Green-Schwarz mechanism involving R-R
axions gives a mass of the order of the string scale to the corresponding gauge boson
\cite{dsw}.  The $SO(8)$ and
$SU(12)$ factors have opposite beta functions, and the latter is strongly coupled in
the IR. Along classically flat  directions, the Chan-Paton group is generically broken
to
$U(1)^4\times SU(2)$. The massless untwisted sector of the closed-string spectrum is
encoded in the following terms of the torus and Klein-bottle partition functions:
\ba {\cal T}_{u} &=& {1 \over 2} \left\{ {| V_2 - S_2 - C_2 |}^2 + \sum_{I=1}^9 \left(
{\left[ \ (O_{2} - S_{2})(\bar{O}_{2} - 
\bar{C}_{2}) \ \right]}_{(I)} + h.c. \right) \right\}
\nonumber \\ {\cal K}_{u} &=& {1 \over 2} \left( V_2 - S_2 - C_2  \right) \quad ,
\label{spectrum}
\ea where $O_2$, $V_2$, $S_2$ and $C_2$ are level-one characters of the transverse
Lorentz group $O(2)$.  The untwisted sector of the ``parent'' type IIB model includes
the 20 NS-NS fields
$(\phi, b_{\mu \nu}, b_{i \bar j}, g_{i \bar j})$ and the 20 R-R fields   
$(\phi^{\prime}, b^{\prime}_{\mu \nu}, b^{\prime}_{i \bar j},  A_{\mu \nu i \bar j})$.
These fill the universal hypermultiplet, as well as 9 additional  hypermultiplets
corresponding to the second cohomology group of untwisted
$(1,1)$ forms, and parametrize the quaternionic manifold
$E_{6(+2)} /(SU(2) \times SU(6))$. This manifold was obtained by c-map 
\cite{cfg} from the special K\"ahler manifold $SU(3,3)/(SU(3) \times SU(3)
\times U(1))$  of the  heterotic string on the $Z$ orbifold \cite{fkp}.  In NS-NS
sector, the open descendant retains the dilaton and a 9-dimensional real slice of the
complex K\"ahler cone corresponding to 
\be Im(J_{i \bar j}) \ = \ Re(g_{i \bar j}) + Im(b_{i \bar j}) \quad .
\label{realslice}
\ee    In the R-R sector, one is left with a mixture of $\phi^{\prime}$ and 
$b^{\prime}_{\mu \nu}$, as well as with mixtures of the other R-R fields.  Though
somewhat surprising, this result, clearly encoded in the vacuum amplitudes of eq.
(\ref{spectrum}), is also supported by the explicit study of tree-level amplitudes, as
well as by some rather compelling duality arguments.  The 20 scalar fields parametrize
a space  $K_I$  that, on general supersymmetry grounds, is a K\"ahler manifold embedded
in
$E_{6(+2)} /(SU(2) \times SU(6))$ and is  expected to have a number of properties:
\vskip 8pt
\begin{itemize}
\item[(a)] $K_I$ should be an {\it irreducible} hermitian symmetric space $G/H$
\cite{hel} of complex dimension 10. This is motivated by previous experience with
conventional orbifold compactifications, where the reduced holonomy leads to scalar
manifolds that are symmetric spaces
\cite{fp}.  Moreover, simple scaling arguments \cite{wroma2} suggest complete mixings
between the type I  dilaton and the other scalars. This marked difference with respect
to the heterotic string reflects the $T$ duality variance of perturbative open-string
vacua.
\vskip 8pt
\item[(b)] $SU(2) \times SU(6) \supset H \supset O(3) \times O(3)$ and $E_6
\supset G \supset O(3,3) \times R^+$, since both the linearly and the non-linearly
realized symmetries of the scalar manifold should be a subset of those of the
``parent'' type IIB string and should  include those of 
$O(3,3) /(O(3) \times O(3))$, the unconventional real slice of the K\"ahler cone chosen
by the world-sheet orbifold to accommodate the NS-NS fluctuations.  This should be
contrasted with the classical CY moduli space of closed strings, that would accommodate
solely $g_{i \bar{j}}$.
\vskip 8pt
\item[(c)] The complex scalars should be in a representation that reduces to the
$(3,3)+(1,1)$ under the decomposition of $H$ to the $O(3) \times O(3)$ subgroup.   
This corresponds to the real slice of the K\"ahler cone discussed above together  with
the NS-NS dilaton, since the NS-NS antisymmetric tensor $b_{\mu \nu}$ is  projected out
of the unoriented closed spectrum. 
\end{itemize}  

These conditions uniquely select 
$Sp(8,R)/(SU(4) \times U(1))$. It should be appreciated that this truncation is vastly
different from the heterotic one.   The simplicity of the type I model of
\cite{wroma2} reflects itself in the linearly realized symmetry,
$H = SU(4) \times U(1)$, with an $SU(4)$ factor strongly suggestive of  a 12D
interpretation.

\section{ Reduction of $N=8$ Supergravity}

The $N=8$ theory includes 70 scalar fields that parametrize the coset
$E_{7(+7)}/SU(8)$ \cite{cj}.  The whole field content  has a natural decomposition in
terms of $N=2$ supermultiplets, as follows from the branching of 
$SU(8)$ in $SU(2) \times SU(6)$,
\be  \left\{2,2\left({3 \over 2}\right),1\right\} + 6\times 
\left\{{3 \over 2}, 2(1), {1 \over 2}\right\} + 15 \times \left\{1,2\left({1
\over 2}\right),2(0)\right\} + 20 \times \left\{{1 \over 2},2(0)\right\}
\label{8to2} \ \ .
\ee The $N=8$ spectrum thus contains 15 $N=2$  vector multiplets and 10 $N=2$
hypermultiplets. Therefore, any truncation of the $N=8$ theory must have $n_V
\leq 15$ and
$n_H
\leq 10$. The untwisted sector of the $Z$ orbifold is the ``maximal'' consistent
reduction, since $n_H=10$ in type IIB (and thus $n_V=9$ in type IIA). The
$Z_3$-projection in the untwisted sector of the type IIB string amounts to retaining
only $Z_3$-singlets in the decomposition of $SU(8)$ into $SU(3)$ induced by
$8 \rightarrow 4 + 4 \rightarrow 3 + 1 + 3 + 1$. Moreover, in the  decomposition
(\ref{8to2}) the hypermultiplets belong to the antisymmetric three-tensor
representation  of $SU(6)$, that has zero $Z_3$ triality.   Further reduction according
to
$6 \rightarrow 3 + 3$ preserves this property and, as a result, all the $N=2$
hypermultiplets in (\ref{8to2}) survive. Conversely, all the 15 vector multiplets are
projected out, since they have nonzero triality. The original 28 vector fields belong
to the antisymmetric  two-tensor of
$SU(8)$, and decompose according to $28 \rightarrow   6 + 3 + 1 + 3 \times (3 + 3^*)$. 
Therefore, only the graviphoton survives the $Z_3$ projection.  For the sake of
comparison, let us note that in the type IIA massless spectrum the  decomposition $8
\rightarrow 4 + 4^* \rightarrow 3 + 1 + 3^* + 1$ implies
$28\rightarrow  8+1 +3 \times (3) +3 \times (3^*)+ 1$. As a result, there are 9
$Z_3$-singlet vectors (together with their $N=2$  superpartners) and the graviphoton.

The truncation that leads to the $N=1$ heterotic string follows from the intermediate
branching of $SU(8)$ in $SU(4)\times SU(4)$. Accordingly, the 
$N=8$ supergravity is to be decomposed in terms of $N=4$ supermultiplets:
\be
\left\{2,4 \left({3\over 2}\right),6 (1),4\left({1\over 2}\right),2(0)
\right\} +4\times\left\{{3\over 2},4(1),(6+1) \left({1\over 2}\right),8(0)\right\}
+6\times\left\{1,4\left({1\over 2}\right),6(0)\right\}
\quad ,
\label{hetdec}
\ee     where the first and last terms correspond to the 
$N=1$ supergravity multiplet in $D=10$, while the middle term accounts for the
additional 10D gravitino multiplet. The 36 scalars in the vector multiplets parametrize
the coset 
$O(6,6)/(O(6) \times O(6))$,  while the two scalars of the supergravity multiplet
parametrize $SU(1,1)/U(1)$ \cite{csf}. Keeping $Z_3$-singlets in the $N=4$ theory 
according to
$6\rightarrow 3+{3^*}$ results in 10 chiral multiplets, one from the supergravity
multiplet and 9 from the vector multiplets. The latter parametrize the special K\"ahler
manifold $SU(3,3)/(SU(3)\times SU(3)\times U(1))$, and  the appearance of $SU(3)$
reflects the holonomy of CY threefolds.

Returning to the open descendant of the type IIB string in \cite{wroma2}, let us
consider the decomposition $8_s \rightarrow 6 +1_+ +1_-$ of the spinor of $O(8)$ in
representations of $O(2) \times O(6)$ induced by the branching $8_v
\rightarrow 4 + 4^*$ of tangent vectors on manifolds of $SU(4)$ holonomy.  As already
noted, the $N=8$ vector fields  belong to the antisymmetric two-tensor of $O(8)$ that
decomposes according to
$[8 \times 8]_A = 15 +6_+ +6_- +1$, giving precisely  one $O(6)$ singlet, the 
graviphoton. This field, however, is projected out of the unoriented closed spectrum
together with one of the
$O(6)$-singlet gravitini. On the other hand, the 40 scalars in the $(2,20)$ of
$O(2)
\times O(6)$  transform as a $10_+$ and a
$10^*_-$, and the projection retains only the
$10_+$ in the unoriented spectrum.  These states belong to the  symmetric two-tensor of
$SU(4)$, or equivalently to the self-dual antisymmetric three-tensor of
$O(6)$, consistently with a scalar manifold $Sp(8,R)/(SU(4) \times U(1))$.

The generalized Siegel upper-half plane
$Sp(2n,R)/U(n)$, familiar from the theory of Riemann surfaces \cite{mum},  may be seen
as the open submanifold ${\cal S}_n$ of complex symmetric
$n\times n$ matrices $\Omega$, with $Im\Omega >0$ \cite{hel}. It is the natural 
extension of the type  IIB
$(\phi,\phi^{\prime})$ system that lives in the upper-half plane.   In this case, the
counterpart of the $SL(2,Z)$ group of $U$ dualities \cite{ht} is $Sp(8,Z)$, that
clearly combines $S$ and $T$ transformations.  As already remarked, this is consistent 
with the properties of perturbative type I vacua, that lack both $T$ and $S$ duality
invariances. On the other hand, one  would expect manifest invariance under dualities
involving only complex structure  deformations \cite{torus} that, however, are  lacking
in the model of \cite{wroma2}. The
$Sp(2n,R)$ group of real matrices
\be S = \pmatrix{A & B \cr C & D \cr}
\ee such that $A^T C = C^T A$, $B^T D = D^T B$ and $A^T D - C^T B=1$ acts projectively
on $\Omega$:
\be \Omega \rightarrow (A\Omega+B)(C\Omega+D)^{-1} \quad .
\ee  In particular, the  continuous Peccei-Quinn symmetry of the R-R fields corresponds
to the  triangular subgroup
\be S_{PQ} = \pmatrix{1 & B \cr 0 & 1 \cr} \quad ,
\ee with $B=B^T$. It is natural to expect that (world-sheet and space-time) instanton
effects induce non-trivial monodromies with respect to 
$Sp(2n,Z)$, so that the moduli space is actually ${\cal S}_n/Sp(2n,Z)$, a
generalization of the fundamental domain of the moduli space of  elliptic curves.  Note
that ${\cal S}_n$ has a natural K\"ahler metric with K\"ahler potential
\be K = - \log \det Im \Omega \quad .
\ee In the model of \cite{wroma2}, a non-perturbatively generated superpotential would
be a modular form of $Sp(8,Z)$.  More generally, for a K\"ahler potential transforming
as $K \rightarrow K - \Lambda_{\Gamma} - \bar{\Lambda}_{\Gamma}$ under the action of a
monodromy group $\Gamma$, the superpotential would be a $\Gamma$-modular form 
transforming as $W \rightarrow e^{\Lambda_\Gamma} \ W$ \cite{flst}.

The analysis may be extended to the untwisted sectors of other abelian  orbifolds. For
the ``parent'' type IIB string, the special K\"ahler manifolds  of the vector
multiplets and the quaternionic manifolds  of the hypermultiplets have been discussed in
\cite{fp}. Open descendants typically  associate to type IIB vector multiplets type I
chiral multiplets, with the same scalar field content, and in these cases the special
K\"ahler manifold is  unaffected by the projection. On the other hand, the
hypermultiplets are truncated to chiral multiplets,  and the number of scalar fields is
correspondingly halved. For instance, for the $Z_2 \times Z_2$ orbifold the
correspondence between the scalar manifolds is 
\be {O(4,4)\over O(4)\times O(4)} \ \rightarrow \ {SU(2,2)\over SU(2)\times SU(2)\times
U(1)} \quad .
\ee
\vskip 24pt
\section{F-Theory Interpretation} 

The compactification of F-theory on the simplest CY fourfold,  the $T_8$ torus,  gives
$N=8$ supergravity in
$D=4$. If $T_8$ is equipped with a complex structure, its fourth cohomology group
decomposes as
\be  H^4 = H^{(4,0)}+H^{(0,4)}+H^{(3,1)}+H^{(1,3)}+H^{(2,2)} \quad ,
\ee  with Hodge numbers $h^{(4,0)}=h^{(0,4)}=1$, $h^{(3,1)}=h^{(1,3)}=16$, 
$h^{(2,2)}=36$.  According to \cite{cj}, the 70 scalar fields  should be assigned to the
$35_s + 35_+$ of $O(8)$, where $35_s$ are metric deformations at fixed volume. Using
$O(8)$ triality 
\cite{cj} one can trade the $35_s$ for the $35_-$. All the 70 scalars 
$\phi_{ijkl}$ can then be associated to the compactified four-form $A_4$, and split
into $\phi_{ijkl}^+$ and $\phi_{ijkl}^-$ ($35_+$ and $35_-$ of O(8)).  The 12D spinors
decompose as $(4,8)$ of $O(3,1)\times O(8)$, where
$O(8)$ is naturally interpreted as the rotation group of the eight compactified
dimensions. This would require 32-component Weyl spinors in $D=12$ with signature
$(11,1)$, and in principle would allow one to obtain both type IIA and type IIB in
$D=10$. On the other hand, a single Majorana-Weyl spinor in $D=12$  with signature
$(10,2)$ would give at most two Majorana-Weyl spinors of opposite chirality in
$D=10$, suitable for type IIA but not for type IIB \cite{bars}.   Dirichlet 3-brane
considerations also seem to indicate an
$(11,1)$ signature \cite{jr} for the putative 12D F-theory, although at this stage an
alternative formulation with $(10,2)$ signature may well be possible. The vector fields
$V_{\mu ij}$ are naturally associated to 
$A^{(3)}_{\mu i j}$.  For the sake of comparison, in M-theory compactified on
$T_7$, 28 scalar fields arise from $g_{ab}$, 7 from
$A^{(3)}_{\mu \nu a}$ and 35 from $A^{(3)}_{abc}$, while 7  vectors arise from
$g_{\mu a}$ and 21 from $A^{(3)}_{\mu ab}$.  The geometrical coupling of \cite{fms}
\be
\int A_4 \wedge F_4 \wedge F_4 \quad  
\ee   induces the $D=4$ couplings 
\be
\int \epsilon^{ijklmnpq} \ \phi^{+}_{ijkl} \ F_{mn} \wedge F_{pq} \quad ,
\ee  where $F_{mn} = d V_{mn}$, indeed present in $N=8$ supergravity
\cite{cj,dwn}.

For M-theory compactified on 8-manifolds, a similar analysis was  performed in
\cite{bb}.  A further reduction of F-theory on $S_1$ allows a comparison with M-theory 
compactified on $T_8$. After suitable duality transformations, the latter contains 128
scalars that belong to the left spinor representation of
$O(16)$ and parametrize the coset 
$E_{8(+8)}/O(16)$ \cite{ms}. The left spinor of $O(16)$ decomposes under $O(8)$
according to $128 \rightarrow 2(1) + 2(28) + 35^+ + 35^-$. This is consistent with the
4D F-theory reduction where, as we have seen, the degrees of freedom can be assigned to
the 4D graviton, to 28 vectors arising from  the three form $A_3$ with two indices
tangent to $T_8$,  and to a completely compactified four-form $A_4$.

The F-theory interpretation of some 6D type I vacua and their relation to
non-perturbative heterotic vacua, proposed in \cite{mv}, has received further support in
\cite{pol}. More general classes of 6D type I models with no tensor multiplets at all
\cite{noid6} correspond to non-trivial K3 fibrations \cite{mv}.

We can only give some hints on the relation of the open descendants of the type IIB
theory  on the  $Z$ orbifold \cite{wroma2} to the F-theory reduction on a CY fourfold.
A generic CY fourfold has
$SU(4)$ holonomy and leads naturally to models with $N=1$ supersymmetry
\cite{cy4}. This leaves a single 4D gravitino,  with a breaking pattern identical to
the one induced by the Klein-bottle projection in the type I model. We have already
seen that, in the unoriented closed spectrum of \cite{wroma2}, the 37 hypermultiplets
turn into 37 chiral multiplets. These allow an interpretation in terms of a fourfold
with
$h^{(1,1)} \neq 0$ and $h^{(1,2)} =0$. The topological couplings 
\be
\int_{CY} A_{i\bar jl \bar m} \ F_{p \bar q} \ F_{s \bar t } \ \epsilon^{ilps}
\epsilon^{\bar j \bar m \bar q \bar t}
\ee   between states associated to the $H^{(1,1)}$ and $H^{(2,2)}$ cohomology groups of
the fourfold\footnote{These couplings were also discussed with  R. Minasian.} give rise
to
$N=1$ axion couplings. An explicit analysis of these couplings may  clarify the role of
the $H^{(2,2)}$ cohomology.

\vskip 24pt
\section{Remarks on Mirror Symmetry and the c-Map}

In the type I model of \cite{wroma2}, the Peccei-Quinn symmetries of the ten real R-R
scalars have a natural geometrical interpretation.  We have seen that the untwisted
moduli space
$Sp(8,R)/U(4)$ is the  generalized Siegel upper half plane, $Im\Omega>0$. The elements
of 
$Re\Omega$ are R-R scalars, while the elements of
$Im\Omega$ parametrize a real slice of the complex K\"ahler cone of the CY threefold.
This actually suggests that open descendants of type IIB compactifications on generic 
CY threefolds involve a new complexification of the classical real moduli space:
\be J \rightarrow J_C^o = i Im(J) + B_{R} \quad ,
\label{complI}
\ee  where, as in eq. (\ref{realslice}), $ImJ$ is the imaginary part of the
complexified  K\"ahler class that survives the world-sheet orbifold projection, and 
$B_R$ is a mixture of R-R fields. Quite differently, in  the heterotic string
\be J \rightarrow J_C^h = ig + B_{NS} \quad .
\label{complhet}
\ee  Since the world-sheet projection of type IIB theories on CY threefolds  relates
$N=1$ to $N=2$ models, this complexification relates chiral  multiplets to
hypermultiplets, giving a new kind of c-map (or, more precisely, an $s_n^{-1}$-map). The
$s_n$-map defined in
\cite{cfg} relates a special K\"ahler manifold
\cite{special} of complex dimension $n$ to a quaternionic manifold of quaternionic
dimension $(n+1)$. In more physical terms, the K\"ahler manifold
$K_n \times SU(1,1)/U(1)$ of $2(n+1)$ NS-NS scalars is ``doubled'' into
$Q_{n+1}$ after the addition of
$2(n+1)$ R-R scalars. The new map for open descendants, that we shall term o-map,  
associates $Q_{n+1}$ to a new K\"ahler manifold $K_I$ for $(n+1)$ NS-NS and
$(n+1)$ R-R real scalars.  It would be interesting to analyze the intrinsic properties
of the real K\"ahler geometry determined by the o-map, since it should be closely
related to properties of the moduli spaces of (CY) fourfolds. Composing the o-map with
the s-map, one would establish a direct correspondence between heterotic and type I 
vacua with $N=1$ supersymmetry.

The complexification of the K\"ahler class of classical CY threefolds is at the heart
of mirror symmetry \cite{mirror}, that should also have a new realization in this
setting, probably related to a 12D interpretation. In this respect, it is worth
recalling that in closed string theories  the complexification of the K\"ahler  cone $J
\rightarrow ig+B_{NS}$ naturally implies that the  classical special geometry of the
$N=2$ moduli space receives quantum corrections from   string (one-brane) world-sheet
instantons. The form of eq. (\ref{complI}) suggests that in this case the $N=1$
classical K\"ahler manifold should receive quantum  corrections from D-brane 
\cite{polchinski} world-volume instantons \cite{bbs}.

\vfill\eject

\begin{flushleft} {\large \bf Acknowledgments}
\end{flushleft} It is a pleasure to thank C. Angelantonj, A.C. Cadavid, R. Minasian and
R. Stora for interesting discussions. A.S. would like to thank the Theory Division of
CERN for the kind hospitality while this work was in progress. The work of S.F. was
supported in part by DOE grant DE-FG03-91ER40662, Task C., by EEC Science Program 
SC1$^*$CT92-0789 and by INFN. The work of the other authors was supported in part by
EEC Grant CHRX-CT93-0340.

\end{document}